\documentclass[10pt,english,oneside,twocolumn,a4paper]{article}
\usepackage{graphicx}
\usepackage[utf8]{inputenc}
\usepackage{times}
\fontfamily{ptm}\selectfont
\usepackage{tabularx}
\usepackage{ragged2e}
\usepackage[singlelinecheck=false]{caption}
\usepackage[T1]{fontenc}
\usepackage[numbers]{natbib}
\usepackage{amsmath}

\usepackage{amssymb}
\PassOptionsToPackage{square,numbers}{natbib}
\usepackage{caption}
\usepackage{subcaption}
\def\url#1{}

\usepackage{babel}
\RequirePackage[blocks]{authblk}
\makeatletter
\let\ps@plain\ps@empty
\def\@xivpt{14pt}
\def\@sect#1#2#3#4#5#6[#7]#8{%
  \ifnum #2<2
    \null\par\vskip-15pt
  \fi
  \ifnum #2>\c@secnumdepth 
    \let\@svsec\@empty
  \else
    \refstepcounter{#1}%
    \protected@edef\@svsec{%
      \ifnum #2<4
        \hb@xt@10mm{\csname the#1\endcsname}\relax
      \else
        \hb@xt@12mm{\csname the#1\endcsname}\relax
      \fi}%
  \fi
  \@tempskipa #5\relax
  \ifdim \@tempskipa>\z@
    \begingroup
      #6{%
        \@hangfrom{\hskip #3\relax\@svsec}%
          \interlinepenalty \@M #8\@@par}%
    \endgroup
    \csname #1mark\endcsname{#7}%
    \addcontentsline{toc}{#1}{%
      \ifnum #2>\c@secnumdepth \else  
        \protect\numberline{\csname the#1\endcsname}%
      \fi 
      #7}%
  \else
    \def\@svsechd{%
      #6{\hskip #3\relax
      \@svsec #8}%
      \csname #1mark\endcsname{#7}%
      \addcontentsline{toc}{#1}{%
        \ifnum #2>\c@secnumdepth \else
          \protect\numberline{\csname the#1\endcsname}%
        \fi
        #7}}%
  \fi
  \@xsect{#5}}
\renewcommand\LARGE{\@setfontsize\LARGE{16}{20}}
\def\abstract#1{\def\@abstract{#1}}
\def\abstractEn#1{\def\@abstractEn{#1}}
\def\titleEn#1{\def\@titleEn{#1}}
\def\@maketitle{%
  \newpage
  \null
  \let \footnote \thanks
    {\LARGE\bfseries\RaggedRight \@title \par}%
    \vskip 1\baselineskip%
    {\normalsize
      \@author\par}%
    \vskip 2\baselineskip%
    \vskip \baselineskip%
    {\section*{Abstract}
      \@abstract}%
  \par
  \vskip 3\baselineskip}

\renewcommand\section{\@startsection {section}{1}{\z@}%
                                   {-3.5ex \@plus -1ex \@minus -.2ex}%
                                   {\baselineskip}%
                                   {\normalfont\Large\bfseries\RaggedRight}}
\renewcommand\subsection{\@startsection{subsection}{2}{\z@}%
                                     {\baselineskip}%
                                     {1ex}%
                                     {\normalfont\large\bfseries\RaggedRight}}
\renewcommand\subsubsection{\@startsection{subsubsection}{3}{\z@}%
                                     {1\baselineskip}%
                                     {3bp}%
                                     {\normalfont\normalsize\bfseries\RaggedRight}}
\renewcommand\paragraph{\@startsection{paragraph}{4}{\z@}%
                                    {1\baselineskip\@plus1ex \@minus.2ex}%
                                    {3bp}%
                                    {\normalfont\normalsize\RaggedRight}}
\renewcommand\subparagraph{\@startsection{subparagraph}{5}{\parindent}%
                                       {3.25ex \@plus1ex \@minus .2ex}%
                                       {-1em}%
                                      {\normalfont\normalsize\bfseries\RaggedRight}}
\parindent\p@
\parskip\z@
\DeclareCaptionLabelSeparator{enskip}{\enskip}
\makeatother

\usepackage{breqn}

\newcommand{\me}{\mathrm{e}}

\title{Generative Adversarial Networks for Synthesizing InSAR Patches}

\author[a,c]{Philipp Sibler}
\author[b,c]{Yuanyuan Wang}
\author[b]{Stefan Auer}
\author[b,c]{Mohsin Ali}
\author[b,c]{Xiao Xiang Zhu}
\affil[a]{Hensoldt Sensors GmbH, 88090 Immenstaad, Germany}
\affil[b]{Remote Sensing Technology Institute (IMF), German Aerospace Center (DLR), 82234 Wessling, Germany}
\affil[c]{Signal Processing in Earth Observation, Technical University of Munich (TUM), 80333 Munich, Germany}

\abstract{Generative Adversarial Networks (GANs) have been employed with certain success for image translation tasks between optical and real-valued SAR intensity imagery. Applications include aiding interpretability of SAR scenes with their optical counterparts by artificial patch generation and automatic SAR-optical scene matching. The synthesis of artificial complex-valued InSAR image stacks asks for, besides good perceptual quality, more stringent quality metrics like phase noise and phase coherence. This paper provides a signal processing model of generative CNN structures, describes effects influencing those quality metrics and presents a mapping scheme of complex-valued data to given CNN structures based on popular Deep Learning frameworks.}

\begin{document}

\maketitle

\section{Introduction}

For image-to-image translation tasks Generative Adversarial Networks (GANs) as initially proposed by Goodfellow et al. \cite{Goodfellow2014} have seen success in a wide range of applications. In the area of remote sensing, conditional GANs have been employed for image translation tasks between optical and SAR intensity imagery. Applications include scene matching between both imaging modalities \citep{Merkle2017}\citep{Merkle2018}\citep{Hughes2018}, the generation of artifical SAR intensity image patches to improve training of scene classifiers \citep{Marmanis2017} or for aiding interpretation of SAR data using artificially created SAR patches \citep{FuentesReyes2019}.

Fundamentally a GAN is composed of a generative (\emph{G}) and a discriminative (\emph{D}) Convolutional Neural Network. Both networks are trained adversarially, such that the generator \emph{G} synthesizes fake images that the discriminator \emph{D} is trying to discriminate from real images on which both networks are trained. Instead of a random excitation, conditional GANs (cGANs) allow for a condition vector or image at the input of the generator, as it is done in the \texttt{pix2pix} cGAN implementation \citep{Isola2017}. The image translation task then is achieved by presenting the source image as the input condition to the generator, retrieving the translated image at the generator output.

Conditioned synthesis of artificial InSAR patches with GANs would have the potential to generate arbitrary amounts of complex-valued SAR imagery for different input conditionings, such as sensor wavelengths, scene types, or different spatial or temporal observation baselines, using pre-aquired images from actual interferometric SAR sensors as training sets. Direct applications would include the generation of additional training set examples for InSAR classifiers or augmenting small InSAR stacks with additional artificial image patches.

Besides good perceptual quality, the generation of artificial complex-valued SAR images for interferometric applications requires more stringent quality metrics, such as low phase noise and a stable phase coherence between the synthesized images. 

This paper gives a simple InSAR imaging model, provides a Convolutional Neural Network (CNN) signal processing model, and evaluates different handling and mapping strategies of complex-valued tensors in Convolutional Neural Networks. 

A new mapping scheme, we refer to as \emph{Nyquist mapping}, is introduced, that allows the re-use of real-valued CNN implementations for complex-valued tensors without changes. After that, the paper presents experimental results based on simple monofrequent test images, real sensor images and simulated SAR scenes and concludes with current work on loss function tuning for promoting coherence between real and fake, that means synthesized, InSAR image patches.

\section{Simple InSAR Imaging Model}

To end up with a simple monofrequent excitation signal for the CNN signal processing model, the following assumptions to the InSAR imaging model were made: Simple „flat earth“ model, flat topography. Ideal scattering, i.e. constant unity-reflectivity function $f$, i.e. $f\left( r, \lambda, \theta \right) = 1$, with $r(\theta)$ as slant range, $\lambda$ as wavelength and $\theta$ as elevation angle. Flat earth: small changes in $\theta$, leading to a phase $\phi(r)$ changing (almost) linearily, resulting in a constant fringe frequency $\omega_0 = \dot{\phi}(r)$.

Under those assumptions a scene response $\bar{x} \left( r, \lambda, \theta \right)$ is given as

\begin{equation}
\bar{x}\left( r, \lambda, \theta \right) = f \left( r, \lambda, \theta \right) \cdot \me^{j \frac{4\pi}{\lambda}r(\theta)}
\end{equation}

with $\phi(r) = \frac{4\pi}{\lambda}r(\theta)$. Given those simplifications, complex-valued SAR scene images $x(m,n) \in \mathbb{C}^{M \times N}$ containing only one discrete spatial fringe frequency with image spectra $X(j\omega_m, j\omega_n)$, can be retrieved.

\begin{figure}[t!]
\centering
	  \includegraphics[trim=0 0 30 0, width=79.5mm]{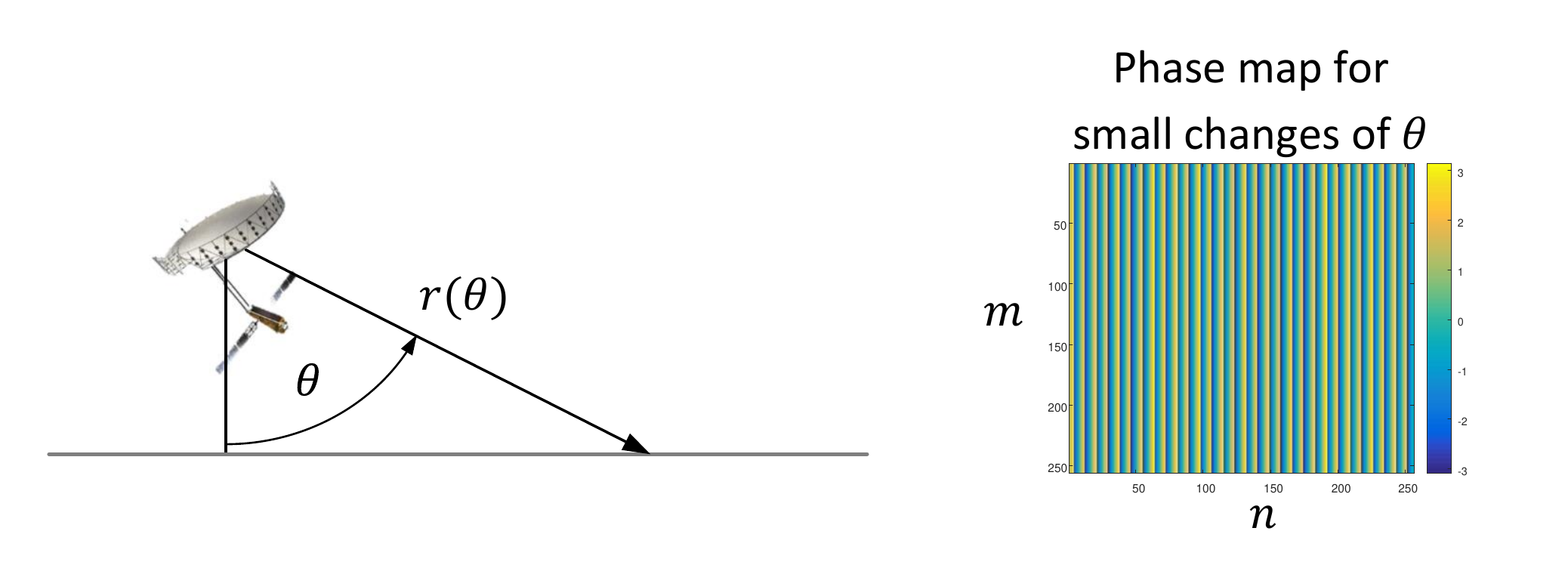}
	  \captionof{figure}{Simple InSAR Imaging Model}
	  \label{fig:imagingmodel}
\end{figure}

\section{CNN Signal Processing Model}

From a system theory perspective two-dimensional Convolutional Neural Networks (CNNs) essentially can be described as several connected layers $l$ of discrete multirate, finite impulse reponse filter banks, mapping the input signal $x^{[l]}_{i}(m,n)$ with $I$ input channels to $J$ output channels. Sets of $I \times J$ trainable weight kernels $w^{[l]}_{ij}(m,n)$ map $I$ input channels to $J$ output channels according

\begin{equation}
z_j^{[l]}(m,n) = \sum_{i} x^{[l]}_{i}(m,n) \ast h^{[l]}_{ij}(m,n) + b^{[l]}_{j},
\label{eq:DiscConv}
\end{equation}

with $h^{[l]}_{ij} = \mathrm{rot180}(w^{[l]}_{ij})$ as two-dimensional impulse responses of the filters and $\ast$ denoting the discrete two-dimensional convolution operation. Within CNN layer $l$ each filter bank is followed by a bias term $b^{[l]}_{j}$, producing an intermediate output $z_j^{[l]}(m,n)$. After that, a nonlinear activation function $a^{[l]}(z)$ generates the output signals 

\begin{equation}
y_j^{[l]}(m,n)=a^{[l]}(z_j^{[l]}(m,n))
\end{equation}

for all $J$ output channels.

The effects of additional support layers like dropout or instance / batch normalization layers, for regularization and improved training convergence, respectively, shall be neglected in this signal processing model for now. 

The model of a CNN as a linear system of layered filter banks with intermediate nonlinearities opens up an interesting explanation for the influences signals are exposed to when passing through the layers of a CNN: On the one hand, for the traversing signal the sets of filter kernels $h^{[l]}_{ij}$ are supposedly trained to be sensitive and selective for certain spatial and spectral features defining the signal. On the other hand, nonlinearities $a^{[l]}$ after each convolutional layer are evoking new spectral content in the form of harmonics and intermodulation products, based on the spectral components that are already present in the signal.

\subsection{Effects of Activation Functions}

For an algebraic model of the effects that nonlinear activation functions $a^{[l]}$ cause to incoming signals, a frequency domain approach shall be discussed: 

To evaluate the absolute positions of all newly generated spectral components, for the moment the \emph{continuous} two-dimensional \emph{Fourier transform} 

\begin{dmath}
\bar{Y}_j^{[l]}(j\omega_m, j\omega_n) = {\iint_{-\infty}^{\infty} a^{[l]}(\bar{z}_j^{[l]}(m,n)) \cdot \mathrm{e}^{-j (m\omega_m + n\omega_n)}\,dm\,dn}
\label{eq:FT}
\end{dmath}

of the continuous output signal $\bar{y}_j^{[l]}(m,n)$ describes the generated spectral content in closed form. Note that in this case the intermediate signal $\bar{z}_j^{[l]}(m,n)$ is assumed to be generated by a two-dimensional convolution integral instead of the discrete convolution operation as given in equation \eqref{eq:DiscConv}.

As a CNN deals with discrete signals, all newly generated spectral frequencies $\omega$ extending the Nyquist frequency interval $[-\omega_s^{[l]}/2,+\omega_s^{[l]}/2[$, with $\omega_s^{[l]}$ as the sampling frequency of layer $l$, fold down from higher Nyquist zones and reappear as alias frequencies $\omega_f^{[l]}$ in this spectral range according 

\begin{dmath}
\omega_f^{[l]} = \omega - \omega_s^{[l]} \lfloor\omega / \omega_s^{[l]}\rceil , 
\end{dmath}

the round-to-nearest-integer operation of $\omega / \omega_s^{[l]}$ given by $\lfloor \omega / \omega_s^{[l]} \rceil$. This holds true for both dimensions $m$ and $n$ and their sampling rates $\omega_{s,m}$ and $\omega_{s,n}$. Evaluating the \emph{discrete} two-dimensional \emph{Fourier transform} (DFT) 

\begin{dmath}
Y_j^{[l]}(j\omega_m, j\omega_n) = {\sum_{m=0}^{M-1} \sum_{n=0}^{N-1} a^{[l]}(z_j^{[l]}(m,n)) \cdot \mathrm{e}^{-j (m\omega_m / M + n\omega_n / N)}}
\label{eq:FT}
\end{dmath}

provides the discrete output spectrum $Y_j^{[l]}(j\omega_m, j\omega_n)$ in a numeric way, containing all in-band and alias frequency components in the frequency interval $[0,\omega_s[$.

\begin{figure}[t!]
\centering
	  \includegraphics[trim=30 0 25 0, width=79.5mm]{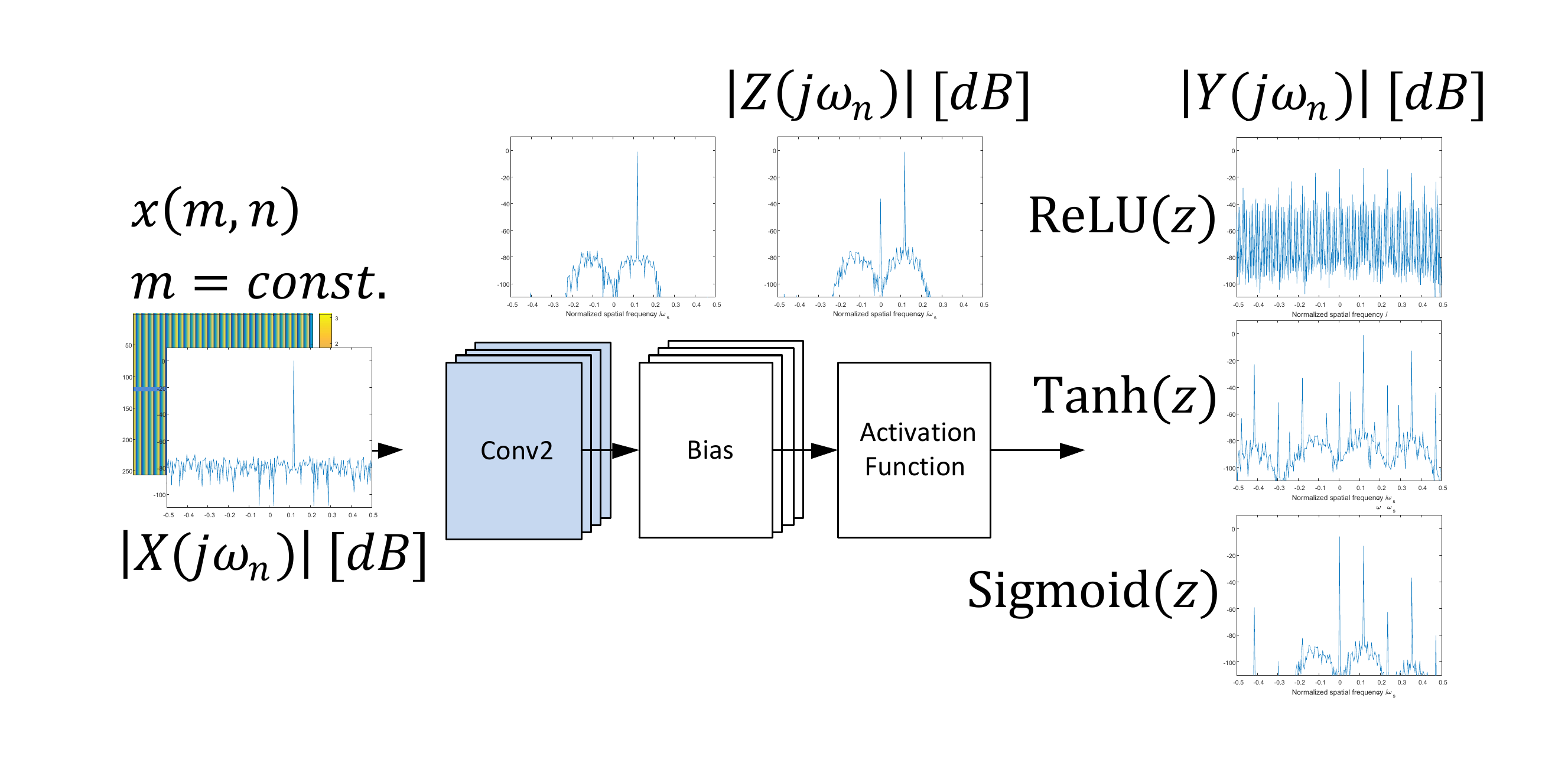}
	  \captionof{figure}{CNN layer with 1-dimensional signal spectra in the frequency domain. Discrete Fourier spectra in log magnitudes are shown.}
	  \label{fig:cnnsys}
\end{figure}

The processing steps of a single CNN layer are presented in \textbf{Figure~\ref{fig:cnnsys}}, from left to right: A (one-dimensional, for clarity, $m = const$) discrete signal spectrum $X(j\omega_n)$, containing only one spatial frequency, as it would be generated from the simple flat earth scene is shown. Following that are the spectra after the convolution block and after biasing $\left(Z(j\omega_n)\right)$ and finally after applying different nonlinear activation functions $\left(Y(j\omega_n)\right)$.

The effect of the nonlinear layer activation functions $a(z)$ to the spectral content of the signals at the output of each layer is quite apparent, with the nonlinearitites causing harmonics and intermodulation between spatial frequencies in the image spectrum. Interestingly, our experiments show that especially the popular ReLU activation function is causing a broad spectral excitation when compared to its Sigmoid and Tanh counterparts. This is even more remarkable as a very sparse input spectrum, containing only one discrete spatial frequency and one bias DC component, is presented to the activation functions.

It shall be mentioned that for \emph{real-valued} input signals $x(m,n) \in \mathbb{R}^{M \times N}$  the effects of a nonlinearity $a(z)$ can be quantified as well in the spatial domain by expanding $a(z)$ with its \emph{Taylor series} around an operating point $z=z_0$:

\begin{equation}
\mathcal{T}\left( a\left(z; z_{0} \right) \right) = \sum_{k=0}^\infty \frac{a^{(k)}\left(z_{0}\right)}{k!} \cdot \left(z - z_{0} \right)^k, \qquad k \in \mathbb{N}, 
\label{eq:TaylorA}
\end{equation}

with $a^{(k)}(z)$ as the $k$-th derivative of $a(z)$. 


The binomial terms $(z-z_0)^k$ are describing the mixing products of the frequency components in $z$ up to the $k$-th order, $a^{(k)}\left(z_{0}\right)/k!$ serves as a scaling factor of those products. Hence, for an intermediate output $z$ containing only one discrete frequency $\omega_0$, a nonlinear activation function $a(z)$ generates an extended spectrum with $k$ harmonic frequencies $k\cdot\omega_0$ according \eqref{eq:TaylorA}. In the Appendix section a real-valued Taylor series expansion for the ReLU activation function is provided.

For forming the Taylor series $a(z)$ needs to be $k$-times differentiable at the operating point. However, it can be shown that most nonlinear activation functions are non-holomorphic in their operating domains \citep{Hirose2013}, that means, in the \emph{complex-valued} case their derivatives $a^{(k)}(z)$ do not exist. Forming a Taylor expansion therefore is not a general option for complex-valued signals.

\subsection{Effects of Resampling}

CNN implementations quite commonly change the spatial resolution of the signals traversing the CNN layers in exchange for modifying the number of filter output channels. Downsampling the signals is usually accompanied by an increase in the number of output filter channels, upsampling the signals by a decrease in the number of output filter channels.

For the generator net G the \texttt{pix2pix} implementation is using an Autoencoder structure with U-Net skip connections: At first, the conditional image is downsampled by factors of 2 to a bottleneck layer with spatial resolution $1\times1$ pixel and 512 filter channels. Out of that abstract latent-space representation the signal is upsampled again by factors of 2 until the final spatial resolution of $256\times256$ pixels at the output layer of G is reached.

However, for both portions of the generator net it can be observed that neither \emph{decimation} filter layers (for downsampling) nor \emph{interpolation} filter layers (for upsampling) are introduced additionally. Spectral aliasing caused by unfiltered resampling therefore is inevitable, with aliasing artifacts further adding to the frequency-synthesizing characteristics of the nonlinear activation functions. In the spatial domain those missing interpolation filter stages in upsampling layers are the cause of "checkerboard" artifacts in generated images \citep{odena2016deconvolution}.

\section{Mapping Strategies for Complex-valued Tensors}

The support for complex-valued tensors and arithmetic is limited or lacking in typical Deep Learning frameworks and therefore in cGAN implementations \citep{Zhu2017} based on them. A direct implementation of a complex-valued Convolutional Neural Network (CV-CNN) is generally requiring Wirtinger calculus in the backward backpropagation step for training \citep{Hirose2013}, as real-valued loss functions and typical activation functions are non-holomorphic, as mentioned in the previous chapter. Those functions are therefore not complex differentiable \citep{Guberman2016}, required gradients for training therefore cannot be established directly. 

\begin{figure*}[t!]
  \centering
  \includegraphics[width=0.75\paperwidth]{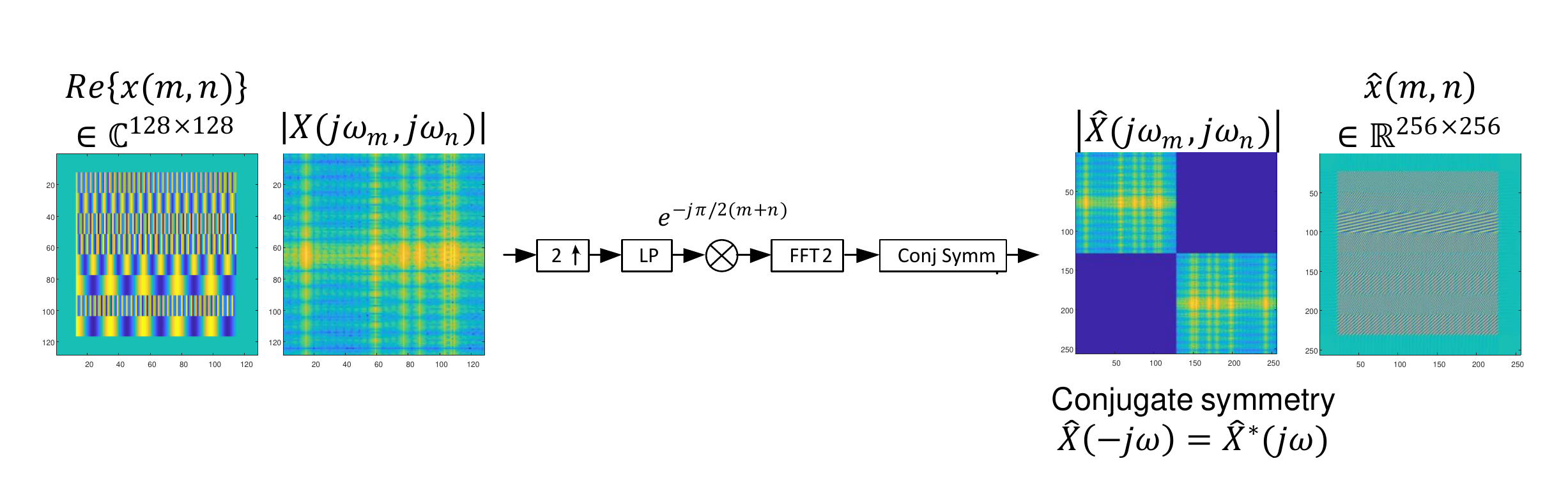}
  \caption{Nyquist mapping scheme for complex-valued tensors, from left to right: complex valued image $x(m,n) \in \mathbb{C}^{M/2 \times N/2}$ (only real part shown), 2D DFT spectrum (log magnitude) of image, processing chain of Nyquist mapping, 2D DFT spectrum (log magnitude) of Nyquist-mapped image with complex-conjugate symmetry, Nyquist-mapped image $\hat{x}(m,n) \in \mathbb{R}^{M \times N}$ in spatial domain. Using Nyquist mapping as proposed in this paper, available real-valued cGAN implementations and Deep Learning frameworks can be reused directly for processing of complex-valued datasets.}\label{fig:nyquistmap}
\end{figure*}

To overcome this limitation, several strategies to map complex-valued data to real-valued-only cGAN implementations were evaluated:

\begin{description}
\item[Direct Real-Imag] Na\"{i}ve approach in direct mapping of real and imaginary components as separated real-valued channels to CNN inputs, mentioned as well in \citep{Hirose2013}\citep{Trabelsi2018}. However, asymmetric Fourier image spectra from complex-valued SAR images suffer from spectral aliasing if their complex components are mapped to separate real-valued channels.
\item[Direct Mag-Phase] Direct mapping of magnitude and phase of complex-valued samples as separated real-valued channels to CNN inputs.
\item[Nyquist mapping] Upsampling by 2, modulating with nomalized frequencies $\omega_m = \omega_n = \pi/2$, FFT2, forcing conjugate symmetry of the Fourier spectra, IFFT2 back to a (now real-valued) spatial domain image. Refer to \textbf{Figure~\ref{fig:nyquistmap}} for details. As the original image spectra are thus centered in the first Nyquist band, the term \emph{Nyquist mapping} was coined for easy reference.

\end{description}

With Nyquist mapping the sampling frequency, and hence the aliasing-free signal bandwidth, is doubled. Therefore the full asymmetric Fourier spectrum of the original complex-valued image $x(m,n)$ can be preserved in the real-valued dataset. 

This advantage comes with one drawback, however: To present Nyquist-mapped, real-valued images to a cGAN like \texttt{pix2pix} with a native resolution of $M \times N = 256 \times 256$ pixels , the original complex-valued images therefore can be only of resolution $M/2 \times N/2 = 128 \times 128$ pixels. This disadvantage can be compensated for by adding additional input and output layers to the \texttt{pix2pix} generator net with a native resolution of $2M \times 2N = 512 \times 512$ pixels. For the experiments presented in this paper this extension to the implementation has not been included yet, however.

\section{Experiments}

Experiments based on the \texttt{pix2pix} implementation \citep{Zhu2017} with three different test sets were performed:

\begin{description}
\item[Onetone] Simple test images $x(m,n) \in \mathbb{C}^{M \times N \times 3}$ with eight random (i.i.d.) monofrequent fringe frequency stripes $\omega_{n}=[-0.5,0.5]\cdot\omega_{s,n}$ (i.e. only in horizontal $n$ dimension, no vertical $m$ frequency component). Input conditioning: 3 channels, $\omega_n$ on channel 0, amplitude on 1, 2 as background channel. Size of training set: 250 images, batch size 3.
\item[OpenSARShip] SLC SAR image patches generated from the OpenSARShip \citep{Huang2018} (Sentinel-1) dataset. Input conditioning: 3 channels showing semantic labels, four classes, background (ocean), ship hull, stern, bow. Size of training set: 250 images, batch size 3.
\item[RaySAR] Random image patches from simulated SLC SAR scene "TUM building" using the RaySAR simulator framework \citep{Auer2016a}. Input conditioning: 3 channels, each channel containing slant range depth information aquired via ray tracing in RaySAR. Size of training set: 500 images, batch size 3.
\end{description}

\begin{figure*}[t]
  \centering
	\begin{minipage}[t]{0.4\paperwidth}
	  \centering
	  \includegraphics[trim=20mm 10mm 0mm 10mm, width=0.4\paperwidth]{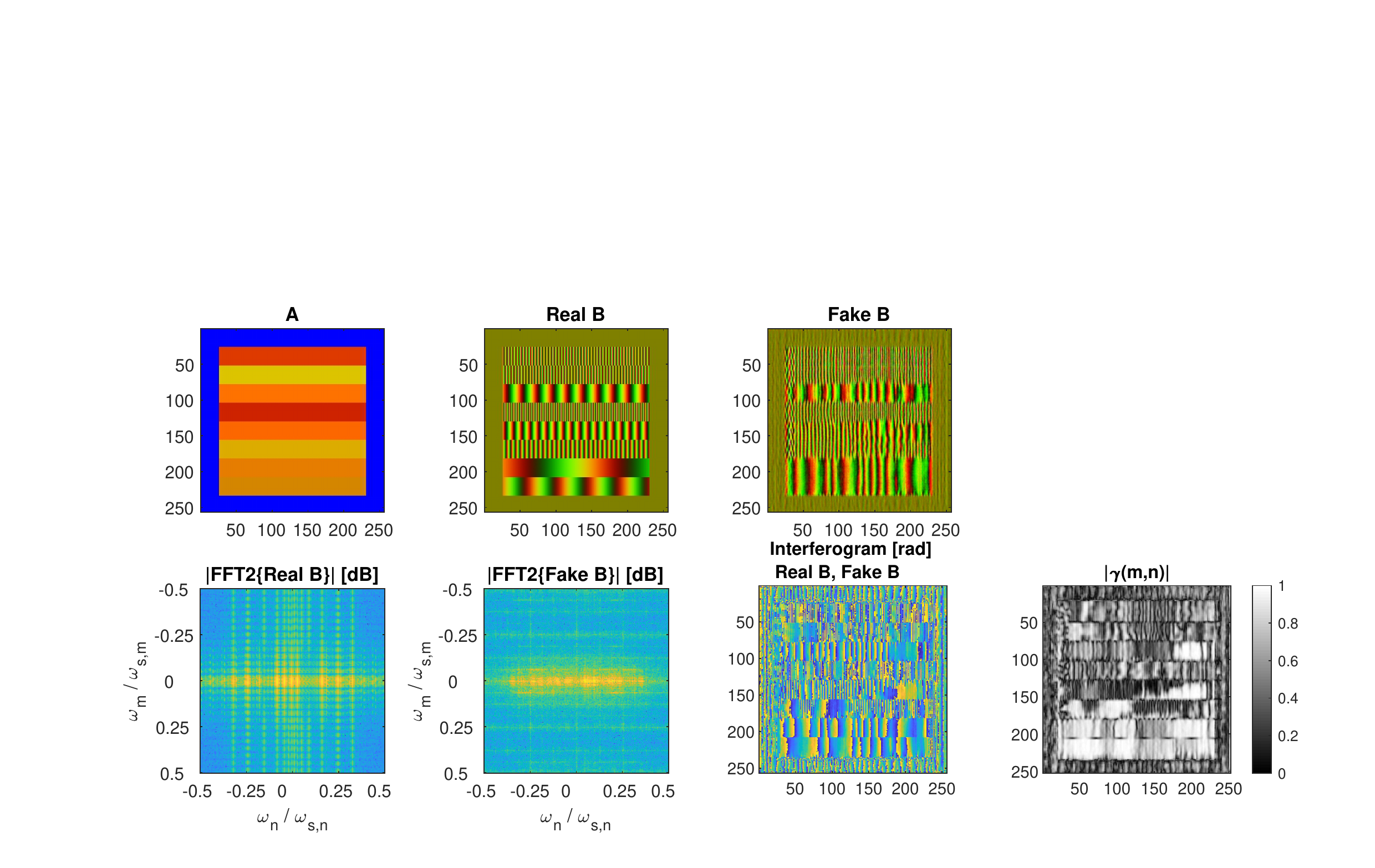}
	  \captionof{figure}{Onetone test image example in \textbf{Direct Real-Imag mapping}, real component on first (red), imaginary component on second (green) channel. Top row, left to right: conditional image, real image, generated fake image. Bottom row: 2D DFT log magnitude spectrum of real image, 2D DFT magnitude spectrum of fake image, interferogram of real and fake, coherence estimate $\mathopen|\tilde{\gamma}\mathclose|$.}
	  \label{fig:onetonereim}
	\end{minipage}
	\begin{minipage}[t]{0.4\paperwidth}
	\centering
	\includegraphics[trim=20mm 10mm 0mm 10mm, width=0.4\paperwidth]{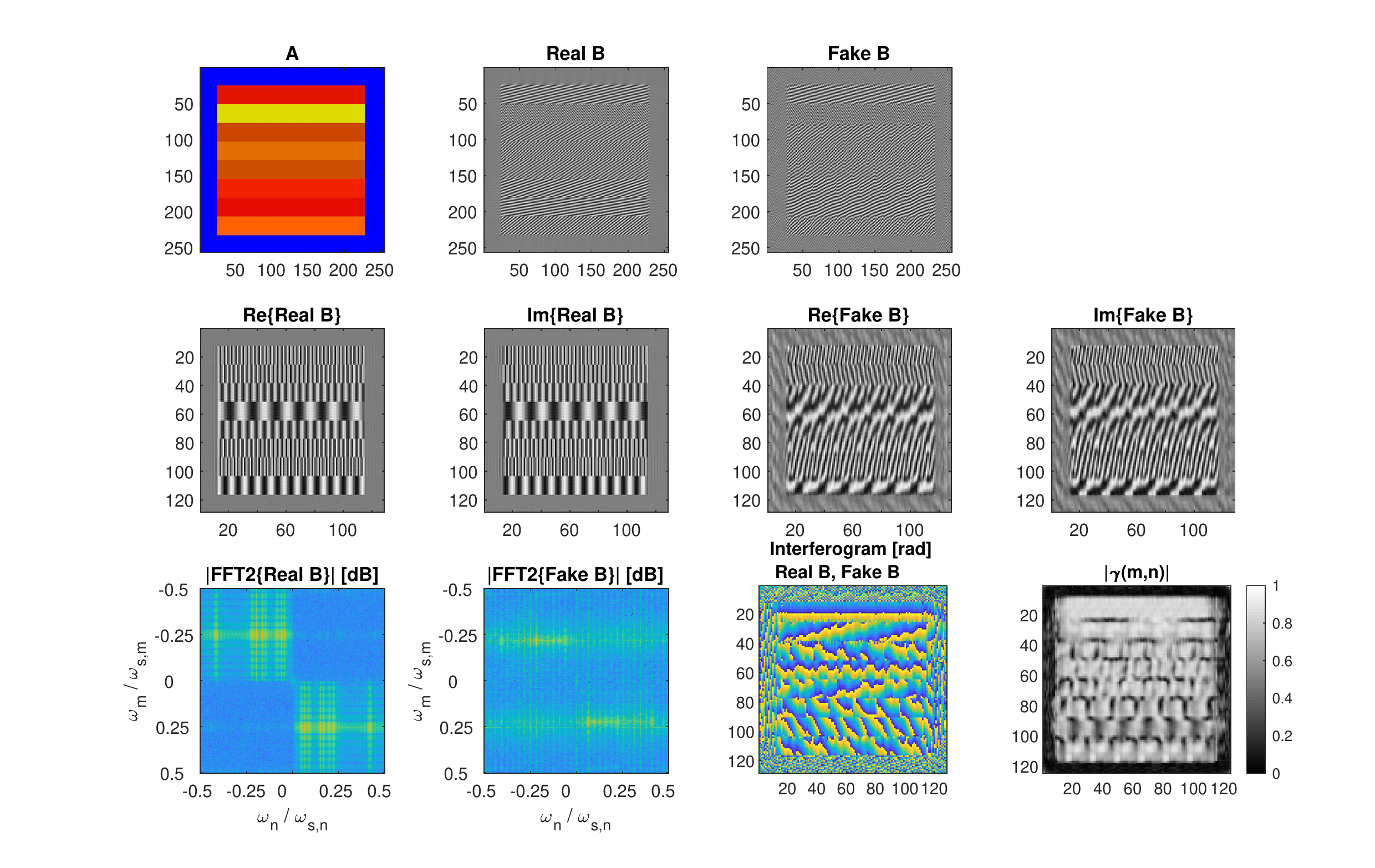}
	  \captionof{figure}{Onetone test image example in \textbf{Nyquist mapping}. Top row, left to right: conditional image, real image, generated fake image in Nyquist mapping. Middle row: Real and imaginary components of decoded real and of fake image, respectively. Bottom row: 2D DFT log magnitude spectrum of real and fake images in Nyquist mapping, interferogram of decoded real and fake images, coherence estimate $\mathopen|\tilde{\gamma}\mathclose|$.}\label{fig:onetoneny}
	\end{minipage}
\end{figure*}

\textbf{Figure~\ref{fig:onetonereim}} presents the \emph{Onetone} training results in Direct Real-Imag mapping for reference. Spectral aliasing is clearly visible in 2D DFT spectrum of fake image due to the separation of real and imaginary components, a low coherence estimate between real and fake (synthesized) image can be observed. 

In comparison the Onetone fake image spectrum generated using a Nyquist mapping training set in \textbf{Figure~\ref{fig:onetoneny}} is showing an improved (asymmetric) spectral estimate of the real spectrum, improved coherence and greater areas with a high coherence magnitude estimate $\mathopen|\tilde{\gamma}\mathclose|$.

\begin{figure*}[t]
  \centering
  \begin{minipage}[t]{0.4\paperwidth}
	  \centering
	  \includegraphics[trim=20mm 10mm 0mm 10mm, width=0.4\paperwidth]{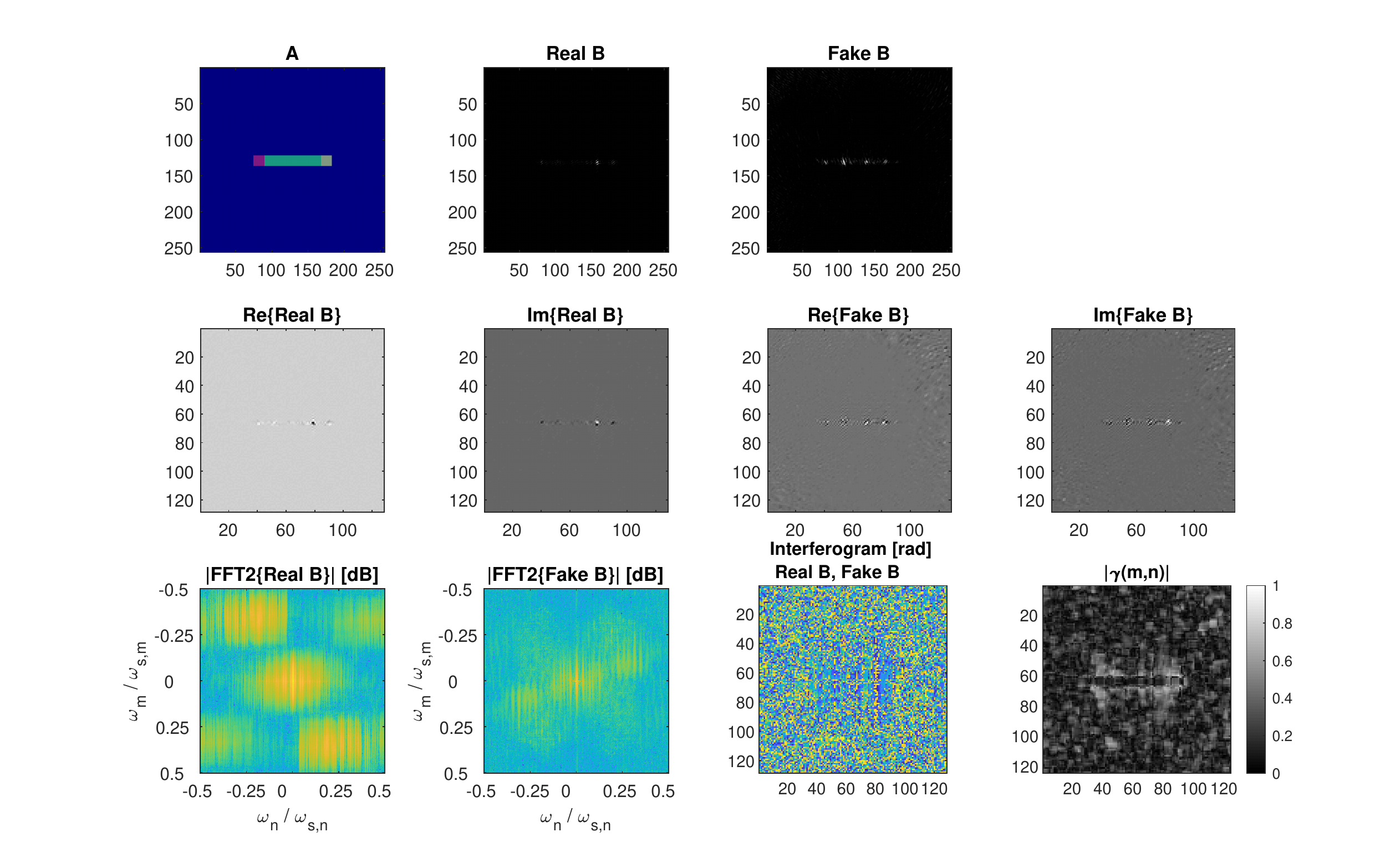}
	  \captionof{figure}{OpenSARShip test image example in \textbf{Nyquist mapping}. Image ordering as in Figure~\ref{fig:onetoneny}.}\label{fig:ossny}  
  \end{minipage}
	\begin{minipage}[t]{0.4\paperwidth}
	  \centering
	  \includegraphics[trim=20mm 10mm 0mm 10mm, width=0.4\paperwidth]{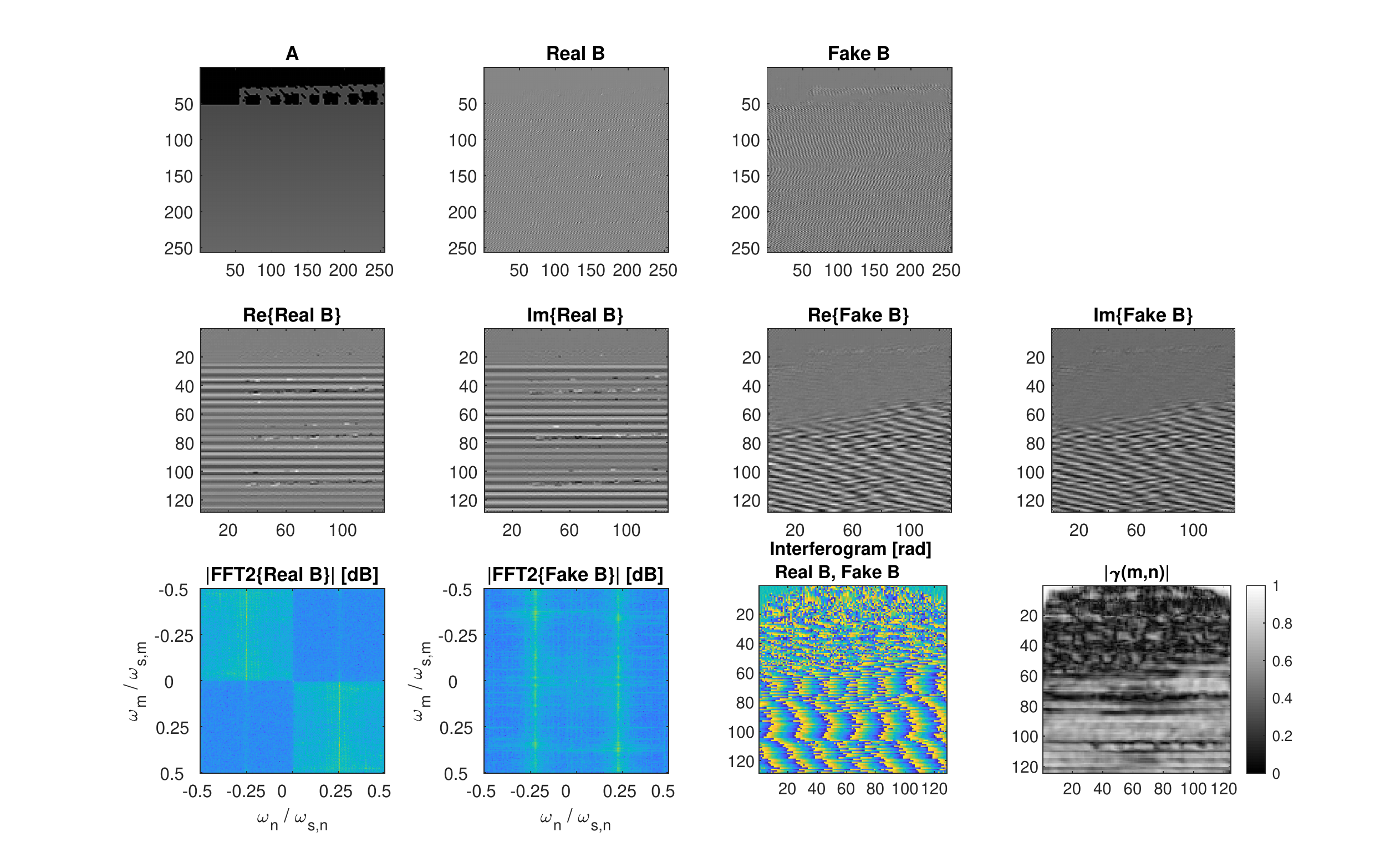}
	  \caption{RaySAR test image example in \textbf{Nyquist mapping}. Image ordering as in Figure~\ref{fig:onetoneny}.}			      \label{fig:raysarny}
  \end{minipage}
\end{figure*}

In the \emph{OpenSARShip} dataset (Nyquist mapping) only small portions (ship hull) of the image patches are showing active regions for SAR interferometry, refer to \textbf{Figure~\ref{fig:ossny}}. Nonetheless, the basic shape is preserved to some extend in the fake image, some local increase of the coherence estimate at the ship's position is visible.

Training the cGAN with artificial \emph{RaySAR} image datasets (Nyquist mapping) and slant range depth images as conditioning provides reproducible test patches for arbitrary scene models. With greater surface regions that actually contain backscatter usable for interferometry, some areas are showing quite a high degree of coherence when generated with Nyquist-mapped training datasets out of RaySAR scenes. Spectral reproduction of the fake images still is lacking, however, also quite some intermodulation noise and upsampling artifacts are present in the fake 2D DFT spectra. 

For both the OpenSARShip and RaySAR datasets training with Direct Real-Imag mapping did not converge to usable results. Direct Mag-Phase mapping did not produce usable results even with the monofrequent Onetone dataset.

\section{Conclusions}

In this paper we presented the possibility of generating artificial complex-valued InSAR datasets using real-valued cGAN implementations. To be able to reuse existing real-valued cGAN implementations and frameworks, complex-to-real mapping schemes for complex-valued tensor data were discussed. Na\"{i}ve mapping schemes were compared to an improved scheme coined \emph{Nyquist mapping}. To understand signal synthesis in a cGAN, a basic signal processing model of CNNs was introduced, describing the effects of spectral filtering in convolutional layers and the effects of nonlinear activation functions and unfiltered resampling that essentially generate new spectral content.

The nonlinear behavior of CNN structures, as cGANs are, still leaves quite room for improvement for InSAR data synthesis in terms of coherence, accuracy in frequency reproduction and purity of the generated spatial frequencies: On the one hand, nonlinearities seem to be the essential "secret sauce" in CNNs to get a full-bandwidth spectral excitation that can be used by the next convolutional filter layer to select and adapt to relevant parts of the spectrum. On the other hand this methodology of repeated nonlinear excitation and subsequent filtering is a noisy form of spectral synthesis, with quite a lot of spurious noise present and still providing only a rough approximation of the desired spectral components and frequencies. 

Current experiments, that are still in progress by writing of this paper, are introducing coherence magnitude loss terms $L_{\mathopen|\tilde{\gamma}\mathclose|}$ to the cGAN generator loss function, which can be formulated in their simplest form as

\begin{equation}
L_{\mathopen|\tilde{\gamma}\mathclose|} = 1 - \mathopen|\tilde{\gamma}\mathclose| .
\end{equation}

\begin{figure*}[t]
  \centering
  \includegraphics[trim=30mm 16mm 30mm 10mm, width=0.5\paperwidth]{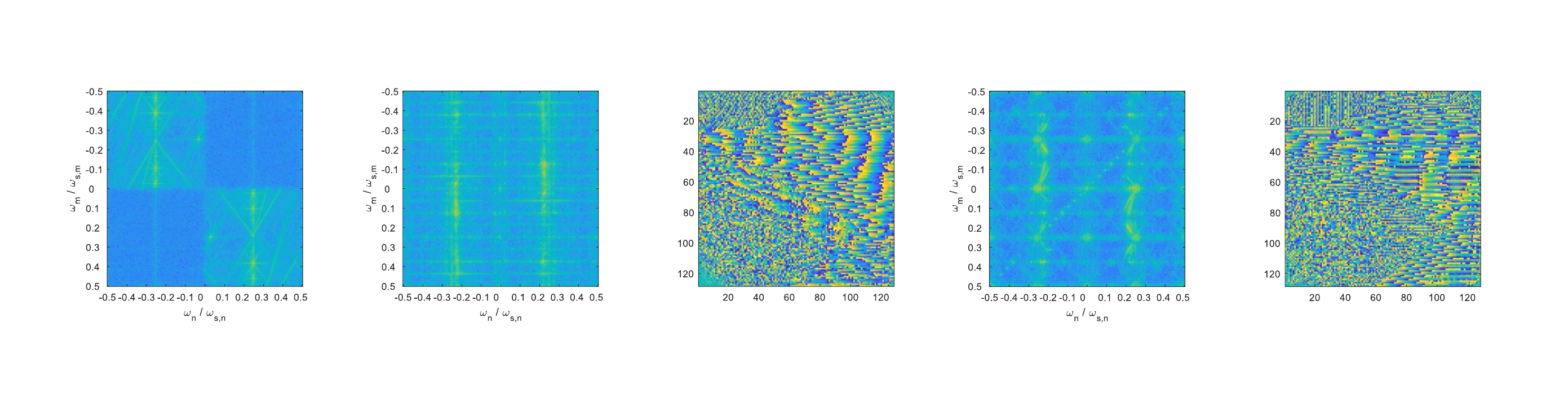}
  \caption{Coherence magnitude loss $L_{\mathopen|\tilde{\gamma}\mathclose|}$ applied to the \texttt{pix2pix} generator net G, loss weight $\lambda_{\mathopen|\tilde{\gamma}\mathclose|} = 1$, influence on RaySAR training dataset. Left to right: 2D DFT log magnitude spectra of real and fake image, cGAN training \emph{without} coherence loss, interferogram of real and fake image. 2D DFT log magnitude spectrum of fake image, training \emph{with} coherence loss $L_{\mathopen|\tilde{\gamma}\mathclose|}$, interferogram. Reproduction of spectral features of the real image is improved in the 2D DFT of fake image, as diagonal spectral features, that extend in both spatial frequency dimensions $(\omega_m, \omega_n)$, start to appear in the fake image image spectrum as well. This effect was only observed so far in generator nets trained with an additional coherence magnitude loss term $\lambda_{\mathopen|\tilde{\gamma}\mathclose|} \cdot L_{\mathopen|\tilde{\gamma}\mathclose|}$.}
  \label{fig:cohloss}
\end{figure*}

Starting to introduce coherence terms to the generator training criteria seems to improve the reproduction of spectral features present in the real training image within 2D DFT spectra of generated fake images, refer to \textbf{Figure~\ref{fig:cohloss}} for initial results. Coherence-inducing loss terms therefore seem likely to be one of the paths to follow for InSAR patch synthesis with cGAN or CNN structures in general.

\bibliographystyle{plainnat}
\bibliography{mybib}\leftskip1mm\advance\labelsep\leftskip

\section{Appendix}

For forming a \emph{real-valued} Taylor series for the ReLU (Rectified Linear Unit) activation function, $a(z) = \mathrm{ReLU}(z)$ needs to be $k$-times differentiable at the operating point, refer to \eqref{eq:TaylorA}. However, the ReLU activation function draws its nonlinearity from a \emph{discontinuity} at $z_{0}=0$, as all negative input values are set to zero and all positive inputs are forwarded undisturbed.

To be able to establish a Taylor series for the ReLU function, this activation function can be expressed in terms of a \emph{continuous}, warped Softplus function with warping factor $\alpha \to \infty$:

\begin{dmath}
\mathrm{ReLU}(z) = \mathrm{max}(0, z) = \lim \limits_{\alpha \to \infty} 
\frac{1}{\alpha} \mathrm{ln}\left(1 + \me^{\alpha z} \right)
\label{eq:ReLUSoftplus}
\end{dmath}

Using the limit approximation for ReLU in \eqref{eq:ReLUSoftplus} the derivatives can be evaluated and a Taylor expansion for the ReLU activation can be formed:

\begin{dmath}
\mathcal{T}\left( \mathrm{ReLU}\left(z; z_{0} \right) \right) = \lim \limits_{\alpha \to \infty} 
\frac{1}{\alpha} \left[\mathrm{ln}\left(1 + \me^{\alpha z_0} \right) + \frac{\alpha^2 \me^{\alpha z_0}}{1!(1+\me^{\alpha z_0})^2} \left(z - z_0 \right) \\ - {\frac{\alpha^3 \me^{\alpha z_0}(\me^{\alpha z_0} - 1)}{2!(1+\me^{\alpha z_0})^3} \left(z - z_0 \right)^2 + \cdots } \right]
\label{eq:ReLUTaylor}
\end{dmath}

\end{document}